\documentclass[twocolumn,trackchanges]{aastex701}
\usepackage{booktabs}

\begin{document}

\title{Observational Tests of Terrestrial Planet Buffering Feedbacks and the Habitable Zone Concept}

\author[orcid=0009-0006-3882-6563,sname='Underwood']{Morgan Underwood}
\affiliation{Department of Earth, Environmental \& Planetary Sciences, Rice University, Houston TX, USA}
\email{morgan.underwood@rice.edu}

\correspondingauthor{Morgan Underwood}
\email{morgan.underwood@rice.edu}  

\author[orcid=0000-0002-5541-9576, sname='Lenardic']{Adrian Lenardic} 
\affiliation{Department of Earth, Environmental \& Planetary Sciences, Rice University, Houston TX, USA}
\email{ajns@rice.edu}

\author[orcid=0000-0002-5293-8872, sname='Seales']{Johnny Seales} 
\affiliation{Los Alamos National Laboratory, Los Alamos New Mexico, USA}
\email{jseales@lanl.gov}

\author{Benjamin Kwait-Gonchar} 
\affiliation{Department of Physics \& Astronomy, Rice University, Houston TX, USA}
\email{bk62@rice.edu}

\collaboration{all}{}

\begin{abstract}

The habitable zone is defined as the orbital region around a star where planetary feedback cycles buffer atmospheric greenhouse gases that, in combination with solar luminosity, maintain surface temperatures suitable for liquid water. Evidence supports the existence of buffering feedbacks on Earth, but whether these same feedbacks are active on other Earth-like planets remains untested, as does the habitable zone hypothesis. While feedbacks are central to the habitable zone concept, one does not guarantee the other--i.e., it is possible that a planet may maintain stable surface conditions at a given solar luminosity without following the predicted $CO_2$ trend across the entire habitable zone. Forthcoming exoplanet observations will provide an opportunity to test both ideas. In anticipation of this and to avoid premature conclusions based on insufficient data, we develop statistical tests to determine how many observations are needed to detect and quantify planetary-scale feedbacks. Our model-agnostic approach assumes only the most generic prediction that holds for any buffering feedback, allowing the observations to constrain feedback behavior. That can then inform next-level questions about what specific physical, chemical, and/or biological feedback processes may be consistent with observational data. We find that [23, 74](95\% CI) observations are required to detect feedback behavior within a given solar luminosity range, depending on the sampling order of planets. These results are from tests using conservative error tolerance--a measure used to capture the risk of false positives. Reducing error tolerance lowers the chance of false positives but requires more observations; increasing it reduces the required sample size but raises uncertainty in estimating population characteristics. We discuss these trade-offs and their implications for testing the habitable zone hypothesis. 

\end{abstract}


\section{Introduction} 

The habitable zone (HZ) hypothesis defines an orbital region around a star where terrestrial planets can maintain liquid water on their surfaces. For our solar system, models place the inner boundary of the zone at 0.99 AU and the outer boundary at 1.70 AU, where 1 AU is the average orbital distance from Earth to the Sun \citep{kopparapuHABITABLEZONESMAINSEQUENCE2013a}. The HZ shifts outward and expands for hotter-burning Sun-like stars and contracts inward for cooler, slow burners like M dwarfs. Its exact bounds depend on assumptions about how atmospheric greenhouse gases respond to varying stellar flux. The inner edge is set by the point at which a runaway greenhouse effect would cause the complete loss of surface water. The outer edge is set by the limit beyond which maximum greenhouse warming can no longer maintain liquid water at the surface \citep{kastingRemoteLifedetectionCriteria2014a}. Classic HZ models typically assume that greenhouse gas buffering is primarily determined by the carbonate-silicate weathering cycle, in which silicate weathering draws down atmospheric $CO_2$ and carbonate precipitation sequesters it \citep{kastingHabitableZonesMain1993a, walkerNegativeFeedbackMechanism1981}. This cycle is thought to stabilize Earth’s climate on geological timescales. 


\begin{figure*}[ht!]
\includegraphics[width=0.9\textwidth]{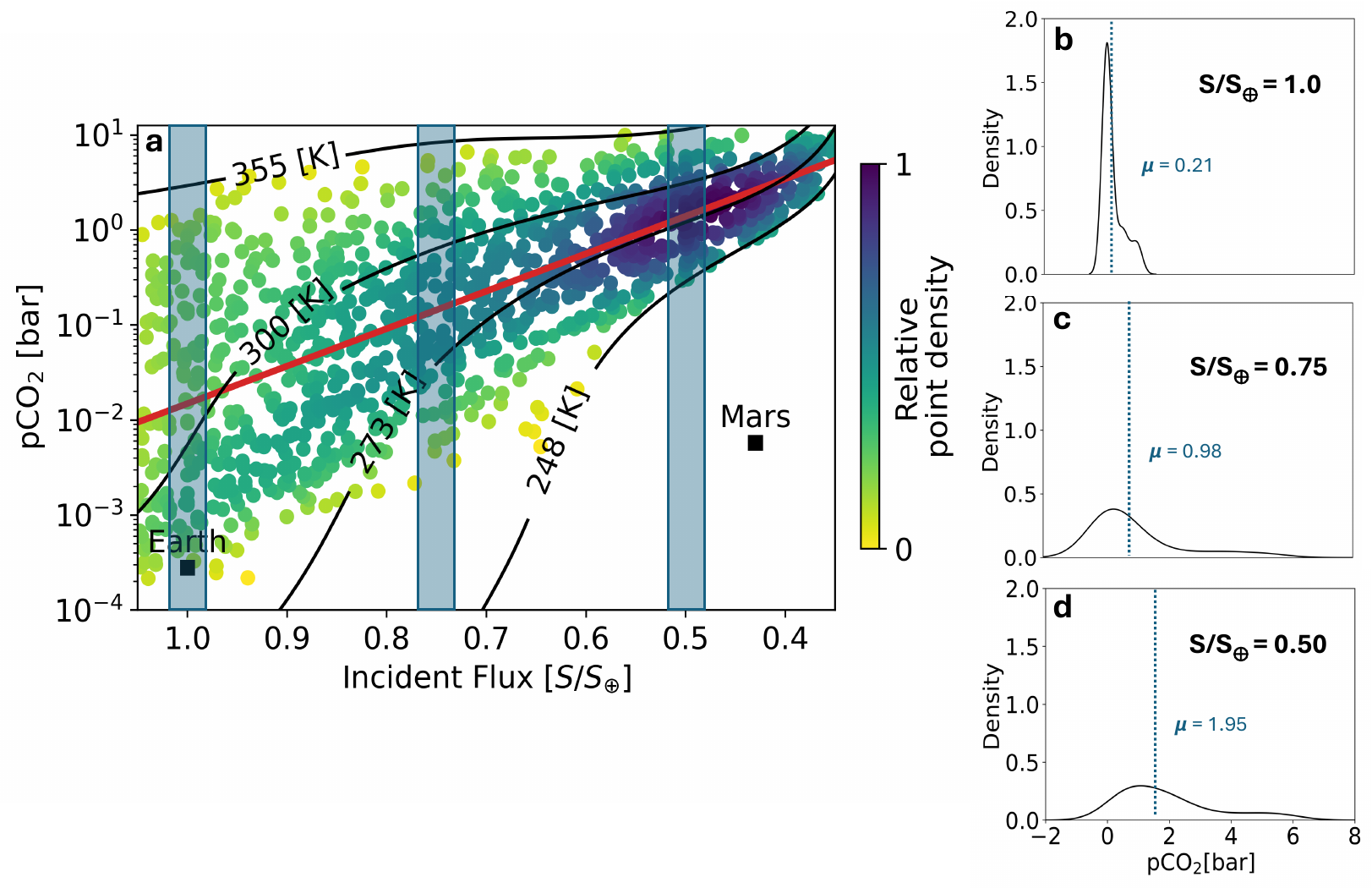}
\caption{\textbf{Carbonate-silicate weathering feedback behavior throughout the habitable zone (HZ).} (a) Reproduction of \cite{lehmerCarbonatesilicateCyclePredictions2020b} results showing stable, Earth-like climates generated by their habitable zone weathering model. The x-axis shows stellar flux normalized to the solar constant ($S/S_\oplus$), the y-axis shows atmospheric $CO_2$ partial pressure ($pCO_2$), and the z-axis shows the distribution of $pCO_2$ values at each $S/S_\oplus$.  Black squares mark modern Earth and Mars (S, $pCO_2$) values.  Point color indicates relative density (yellow = low, purple = high). The model predicts increasing atmospheric $CO_2$ with decreasing stellar flux within the HZ. (b-d) Kernel density estimates (KDE) of $pCO_2$ for simulated planets at 1.0 (b), 0.75 (c), and 0.50 $S/S_\oplus$ (d). While panels b-d reflect model parameter variations in $pCO_2$, similar distributions would be expected in observational data if a $CO_2$-buffering feedback is operative, highlighting the ideal dataset for applying our method.}
\end{figure*}

Model calculations of HZ limits assume that a carbonate-silicate weathering cycle buffers atmospheric $pCO_2$ on planets within the HZ \citep{kastingHabitableZonesMain1993a, kopparapuHABITABLEZONESMAINSEQUENCE2013a, krissansen-tottonConstrainingClimateSensitivity2017, rushbyLongTermPlanetaryHabitability2018a}. Under this assumption, $pCO_2$ adjusts in response to variations in solar flux, as depicted in Figure 1a \citep{lehmerCarbonatesilicateCyclePredictions2020b}. The figure shows the modeled relationship between normalized solar incident flux ($S/S_\oplus$) and steady-state atmospheric $pCO_2$ for Earth-like planets with active carbonate-silicate cycling. Scatter in the predicted greenhouse gas partial pressures reflects parameter uncertainties within the specific model used, while the point densities along the best-fit line indicate that, despite these uncertainties, predicted $pCO_2$ values have a well-defined mean value that varies with solar luminosity. 

The trends plotted in Figure 1a are a result of temperature-dependent feedback cycling, associated with carbonate-silicate weathering, being incorporated into the models used to generate the plot. Figures 1b-1d show how $pCO_2$ values of simulated HZ planets are distributed within certain luminosity windows. As the incident flux ratio decreases from 1 (Earth-like) to 0.5 (Mars-like), the shapes of the probability density functions (PDF) change. These morphological differences are model dependent, reflecting parameter uncertainties specific to the model used. However, the more generic structure of the PDF’s, a single peak with decaying variations from the peak, is characteristic of any negative feedback mechanism \citep{lenardicHabitabilityProcessState2021}. This property enables an alternate testing approach in which observed values of $pCO_2$ can be used to determine PDF shapes. If those observations lack the generic signature of buffering feedback, then a critical aspect of the HZ hypothesis would be refuted. If the signature is present, it can be used to characterize feedback strength, given a large enough sample. This approach offers a model-agnostic pathway to testing for planetary-scale feedback and the HZ hypothesis. 

There are several reasons to adopt model-agnostic approaches to probe exoplanet habitability: 1) the carbonate-silicate weathering cycle is not the only buffering feedback capable of maintaining habitable conditions \citep{lenardicDifferentMoreValue2019a}; 2) Different feedbacks may operate at different locations within the HZ;  3) Detected terrestrial exoplanets in other stars' HZs are diverse \citep{grenfellPossibleAtmosphericDiversity2020, ballmerDiversityExoplanetsInterior2021, rogersMost16Earthradius2015}, making Earth-based assumptions potentially unreliable. Limited observational data further favor testing the most generic of hypothesis predictions. As of now, atmospheric studies with the James Webb Telescope focus on planets orbiting ultracool dwarf stars, of which only a handful host transiting planets in their habitable zones \citep{dewitRoadmapAtmosphericCharacterization2024}. While these observations can inform predictions about the broader population, they will likely accumulate slowly until next-generation telescopes come online. This motivates testing frameworks that leverage limited data while remaining flexible for future data acquisition. 

In this study, we present a framework to estimate the number of observations required to characterize population-level behavior of terrestrial planet atmospheres, accounting for two end-member levels of statistical risk, represented by differing decision thresholds. This can be leveraged to test two ideas: 1) whether any feedback is active within a given luminosity window, and 2) whether the behavior of these feedback(s) is consistent with the HZ hypothesis. 

Although our approach differs from that of \cite{lehmerCarbonatesilicateCyclePredictions2020b}, the methods are complementary. To clarify this, we start by reproducing their results while relaxing an assumption regarding planetary population limits. This demonstrates the robustness of the results and highlights the limits of previous statistical tests, motivating the complementary approach developed here. This forms section 2 of this paper.  Methods are presented in Section 3, results in Section 4, and Section 5 frames our findings within a risk-cost-benefit perspective, considering the impact of observational uncertainties. We also discuss how this framework could be extended to exoplanet data beyond greenhouse gas concentrations, e.g., albedo variations.

\section{Reproducibility Results and Motivation for Expanded Statistical Tests}
\label{sec:style}

\cite{lehmerCarbonatesilicateCyclePredictions2020b} first modeled greenhouse gas concentrations using a coupled climate-carbonate-silicate weathering model, solving for steady-state $pCO_2$ as a function of stellar luminosity. Predicted values, shown as discrete points in Figure 1a, exhibit scatter as a result of uncertainties in model parameters (e.g., internal heat of a planet, land fraction, strength of weathering feedback). A portion of the full parameter space was sampled to construct a population of 1050 model planets, excluding combinations that produced frozen planets or complete loss of surface water. The resulting best-fit trend predicts a log-linear relationship between $pCO_2$ and incident flux. Lehmer et al. proposed that searching for this two-dimensional (S, $pCO_2$) relationship could determine if the predicted trend (informed by the HZ theory) is consistent with observations. Resolving such a trend requires a sufficiently large dataset, so they used a Kolmogorov-Smirnov (KS) test to quantify the number of planets needed to distinguish between two populations, one that is consistent with HZ predictions and one that is not (the null hypothesis). 

\begin{figure*}[ht!]
\centering
\includegraphics[width=0.9\textwidth]{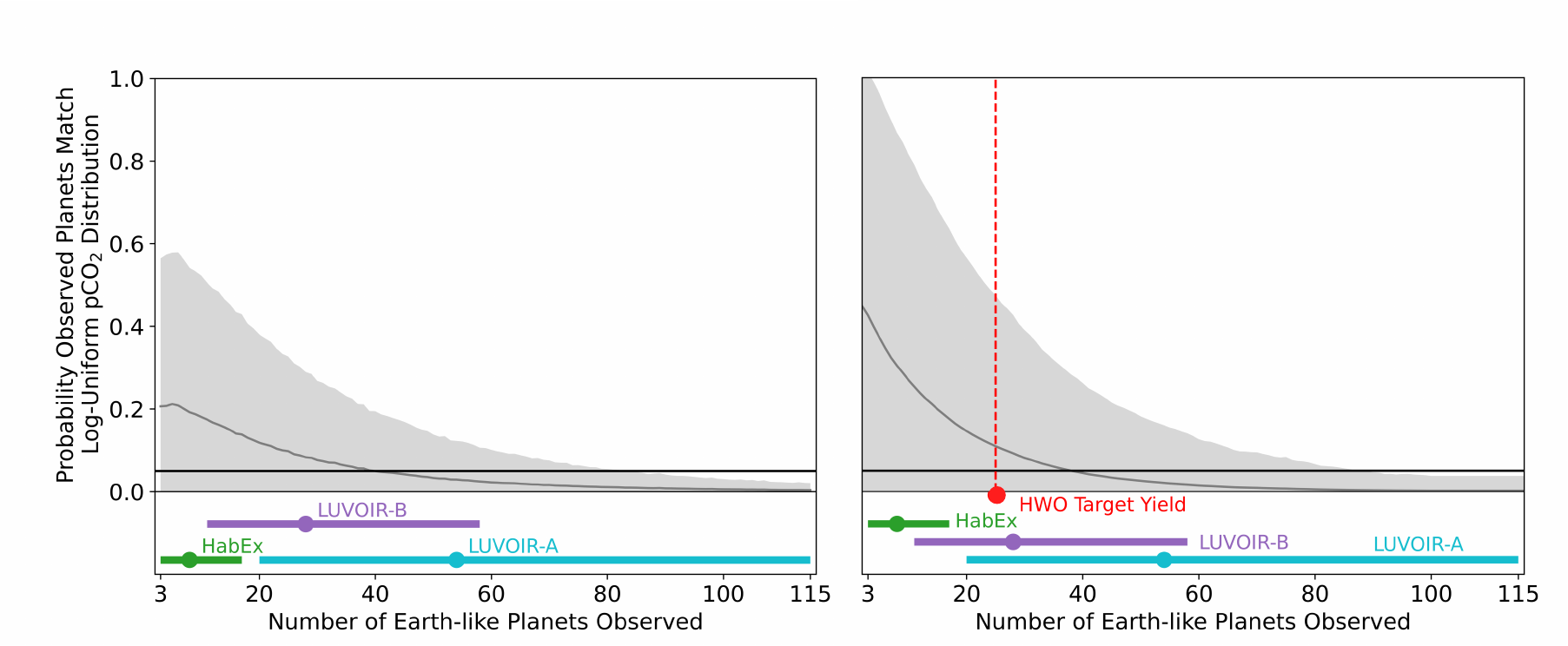}
\caption{\textbf{Reproduction of Lehmer et al.’s results showing probability that exoplanets with active carbonate-silicate weathering cycles could be misidentified as following a log-uniform $pCO_2$ distribution.}  In both panels, the x-axis shows the number of observed Earth-like exoplanets. The solid gray curve represents the mean p-value from Kolmogorov-Smirnov (KS) tests comparing samples of simulated planets with active carbon cycles to a log-uniform distribution (the null hypothesis in this case). The gray shaded region indicates the $2\sigma$ range of p-values. The black horizontal line marks the $5\%$ significance level for rejecting the null. Along the bottom, mission yield estimates are shown for HabEx \citep{gaudiHabitableExoplanetObservatory2020}, LUVOIR-A , and LUVOIR-B, with lines and circles indicating expected Earth-like exoplanet detections. These concepts have since been integrated into the Habitable Worlds Observatory (HWO)  \citep{dressingHabitableWorldsObservatory2024}, NASA’s next astrophysics flagship mission.}
\end{figure*}

Figure 2a shows their results. The x-axis shows the number of observed Earth-like planets, while the y-axis displays the p-values obtained from the KS tests. The p-value represents the probability that the data support the null hypothesis ($H_0$). The significance threshold of $\alpha$ = 0.05 is marked by the horizontal black line. $H_0$ is rejected once the test's p-value falls below this threshold. The objective is to determine the smallest sample size required to reliably distinguish between planetary populations, while also avoiding false trends that might arise from scatter in small datasets. 

We confirmed Lehmer et al.'s results by applying our own framework to their climate-carbonate-silicate-weathering model output. Our approach differs in two main ways: 1) we use a one-dimensional test on $pCO_2$ distributions within specified stellar flux windows, rather than a two-dimensional test in (S, $pCO_2$) space and 2) we incorporate a sequential analysis, testing progressively larger subsets of the data to see how detection thresholds evolve as sample size grows, as opposed to using fixed sample sizes. 

In Lehmer et al.'s framework, $H_0$ states that the (S, $pCO_2$) values for planets with active carbonate-silicate weathering are statistically indistinguishable from a log-uniform distribution--i.e., no systematic trend is present. Figure 2a shows that $H_0$ is eventually rejected, supporting the predicted HZ trend. In our one-dimensional reproduction of Lehmer et al.'s results, $H_0$ instead specifies that the log-normal $pCO_2$ distribution (Fig. 1a) across the HZ is statistically indistinguishable from a log-uniform distribution, defined by the observed data range. The alternative hypothesis, $H_A$, is that the observed $pCO_2$ distribution deviates from a log-uniform, indicating that $pCO_2$ is not best described by this type of distribution. 

In our framework, the sequential analysis applies KS tests iteratively to increasingly larger subsets of the data. This provides insight into the minimum number of observations necessary to reject $H_0$ without requiring known observational constraints. In contrast to Lehmer et al., who evaluate fixed sample sizes, our method sequentially analyzes multiple subsamples increasing in size until certain criteria are met (expanded on in Section 3). As shown in Figure 2b, this adds uncertainty in tests on small datasets but ultimately achieves the same result as Lehmer et al. 

Lehmer et al.'s method is designed to detect a 2-D trend between solar luminosity and model-predicted steady-state $pCO_2$ values. However, the sample size required to confirm that trend may not be adequate to determine the $pCO_2$ distribution within any specific solar luminosity window. It is also possible that early exoplanet observations will not span the full luminosity range of the habitable zone, and instead, planets that are most ``Earth-like" or easiest to observe will be targeted. The question then becomes: what information can be extracted from a more limited sample? Our approach addresses this by focusing on constraining the distribution of $pCO_2$ values and whether it is consistent with the presence of buffering feedbacks – without requiring observations over the wide solar luminosity range needed to test the habitable zone hypothesis. As more observations accumulate, the approach can be extended to test habitable zone predictions. An advantage of our approach is that it allows for the potential that operative feedbacks vary across a habitable region – the models of Lehmer et al., like most habitable zone models, consider one dominant feedback to operate across the habitable zone.

Although the predicted $pCO_2$ PDFs in Figure 1 result from model parameter variations, we would expect similar distributions in the observational data if a buffering feedback is operative within a planetary population, regardless of the model. A buffering feedback does not mean that planetary $pCO_2$ remains at a steady state equilibrium value. Instead, the expectation is that the values can fluctuate, but the feedback keeps them confined to a region around an equilibrium value. This is consistent with paleoclimate observations from the Earth \citep{fosterFutureClimateForcing2017, learGeologicalSocietyLondon2021}. 

\begin{figure*}[ht!]
\centering
\includegraphics[width=0.9\textwidth]{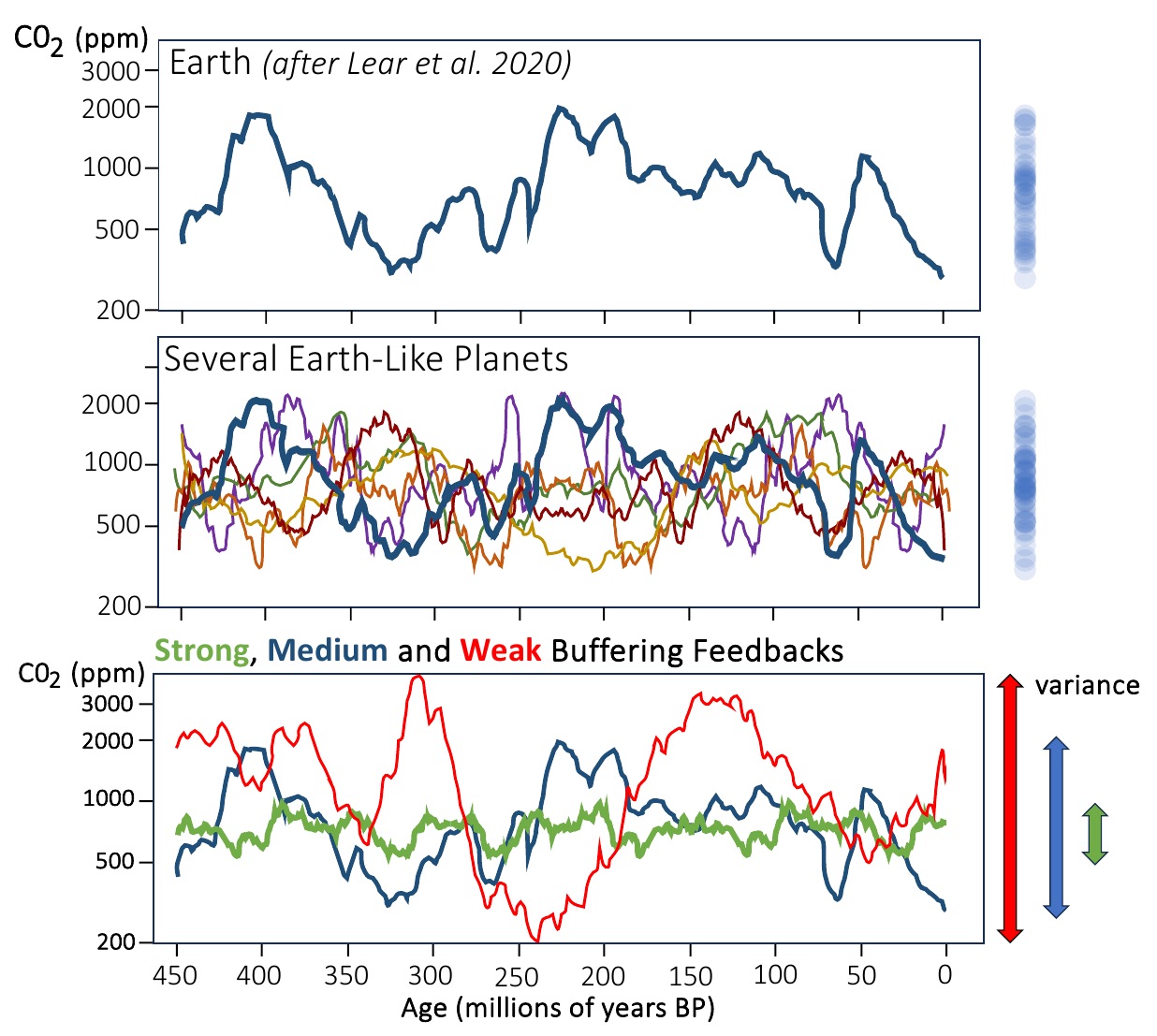}
\caption{\textbf{Schematics of observable $CO_2$ feedback behavior.} The x-axis shows time (in millions of years), and the y-axis shows $pCO_2$ (in ppm). The top panel shows a time series of Earth’s $pCO_2$ for the past 400 million years, based on a smoothed fit to multiple paleoclimate datasets \citep{learGeologicalSocietyLondon2021}. The blue dots to the right are samples from the time series at 50 Myr intervals and illustrate the relationship between a probability density function (PDF) and $pCO_2$ observations. The middle panel presents hypothetical $CO_2$ time series from other Earth-like exoplanets. If similar negative feedbacks modulate their atmospheres, we expect analogous fluctuations. Same as the top panel, the blue dots to the right of the middle panel show samples from the time series at 50 Myr intervals, illustrating sampling planets with similar time series with different fluctuations produces an observable normal distribution. The bottom panel explores how the amplitude of $pCO_2$ fluctuations depends on the strength of the climate feedback(s) at work (red = weak, blue = medium, green = strong). To the right of the bottom panel, we show how this affects the variance of the resulting distributions.}
\end{figure*}

The top panel of Figure 3, adapted from \cite{learGeologicalSocietyLondon2021}, shows a times series of $pCO_2$ plotted against geologic time over an interval where changes in the Sun’s solar luminosity are small enough to be treated as constant. The density plot to the right shows the $pCO_2$ values obtained by sampling the time series at 50My intervals. For a population of Earth-like planets with buffering feedback that modulates greenhouse gas levels, the same $pCO_2$ values are not expected for all planets at the same point in time. Instead, what can be expected is that such planets exhibit $pCO_2$ fluctuations around a mean, with the magnitude of those fluctuations being similar across the population. The middle panel of Figure 3 illustrates this behavior. The starting question for the methodology is: how many observations are required to constrain this distribution well enough to determine whether it reflects the operation of buffering feedbacks among the terrestrial planet population? 

Constraining a distribution can reveal whether a buffering feedback may be operative and provide insights into its strength. For example, weaker feedback produces larger variations, reflected in a wider distribution, as shown schematically in the bottom panel of Figure 3. In the next section, we present a framework to estimate the minimum number of observations needed to resolve a $pCO_2$ distribution for a population of planets. The method is general and can be applied to any observable planetary characteristic (e.g., global albedo), not just greenhouse gas concentrations.  

\section{Methodology} \label{sec:methods}

\begin{figure*}[ht!]
\centering
\includegraphics[width=0.9\textwidth]{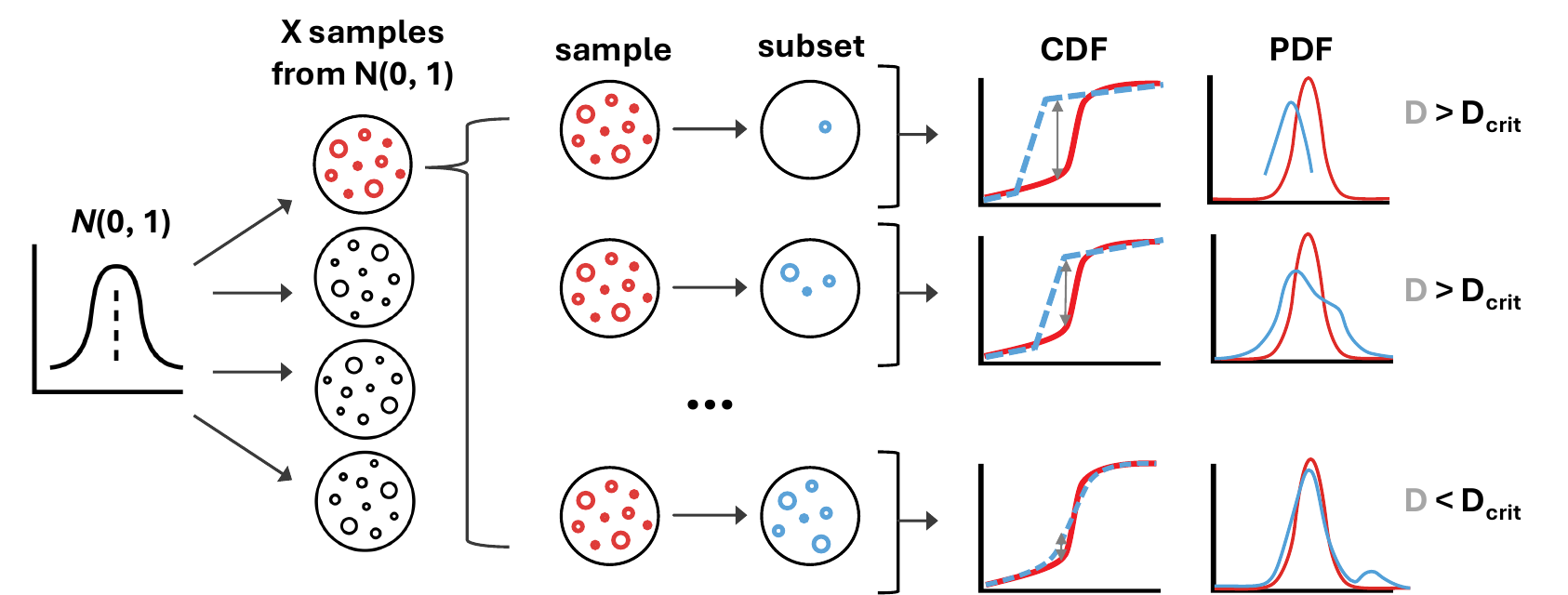}
\caption{\textbf{Workflow for estimating minimum sample size via distribution testing.} A standard normal reference PDF is taken as the true model. From it, X number of samples are drawn, where each sample undergoes a sub-routine. Using the Kolmogorov-Smirnov (KS) test, the sample subset shown in blue is compared against the sample. The KS statistic, D (the largest vertical gap between cumulative distribution functions) is compared with a chosen critical value, $D_{crit}$. The subset size at which D exceeds $D_{crit}$ is recorded and represents the least number of planet observations required to recover the model’s distribution parameters.}
\end{figure*}

In the previous section, we linked planetary-scale feedbacks to probability distributions. For example, if a negative feedback like the carbonate-silicate weathering feedback is at work, its effects will be visible in atmospheric $pCO_2$ samples. The question we address is the minimum sample size needed to approximate the statistical parameters of the population the samples will come from. We use hypothesis testing to determine the sample size necessary to infer the mean and standard deviation of the population, within a pre-defined error tolerance. The following sections describe our approach step-by-step, including the assumptions made about the population, the simulation of the observational process, and the statistical techniques used to recover population parameters from limited data. 

Figure 4 is a visual outline of our framework. We begin by assuming an initial distribution of the expected data. This is analogous to establishing a weak prior in a Bayes framework. For proof of concept, we use a standard normal distribution, N(0, 1), in this study, as the simplest form consistent with a buffering feedback. However, the framework can accommodate any statistical function, making it adaptable to testing different hypotheses. We then draw random samples from N(0,1). Each sample undergoes a subroutine that simulates the telescope observation process, accounting for both sampling error and uncertainty in observation order. As the simulated observations accumulate, a hypothesis test compares the sample to the N(0,1) model. Once the differences fall below a critical threshold, the sample is considered consistent with the model. This process estimates the minimum number of observations needed to recover the model's parameters within a specified error tolerance. We discuss each of these steps in further detail below. 

\subsection{Key Terms}

Before describing sampling procedures, we define key terms used in this study. A \textit{population} is a group of items that share at least one property relevant to a statistical analysis. Here, the population of interest is terrestrial exoplanets and their atmospheric $CO_2$ concentrations at specific points within the HZ. A \textit{population distribution} describes the statistical distribution from which the samples are drawn. A \textit{sample} is a subset of elements randomly drawn from the population, and each \textit{element} is a single member of the population--here, an individual $pCO_2$ observation.

\subsection{Sampling Routine}

The sampling routine involves drawing independent, identically distributed (i.i.d.) random samples from N(0, 1), shown in Step 2 of Figure 4. Monte Carlo sampling is used to capture the inherent variability of the random sampling process \citep{muthenHowUseMonte2002}. Ideally, increasing the sample size, rather than the number of random samples, would reduce sampling error, since larger samples are more likely to accurately reflect the true population. However, in exoplanet studies, acquiring more observations is often constrained by cost and practical limitations. Therefore, our approach evaluates the quality of predictions achievable with a relatively small sample size, while accounting for sampling error, as well as errors related to the sequence observations are made in (discussed in Section 3.3). In this study, we draw 10,000 samples of size N=100 from the N(0, 1) distribution. The sample size can be adjusted in future tests if additional observations become feasible. 

\subsection{Observation Sequencing}

In an observational survey of multiple planets, the number of potential observation sequences introduces uncertainty in determining the requirements needed to test habitability hypotheses \citep{garcia-piquerEfficientSchedulingAstronomical2017}. Collecting many observations would reduce this uncertainty but is costly, while too few observations risk detecting a trend that is not representative of the population. To address this sequence uncertainty, the second level of sampling simulates different observation sequences to estimate the minimum sample size to predict the N(0, 1) model parameters for a range of sequences. This is done by repeatedly drawing random samples from the model, as shown in Step 3 of Figure 4, and stepping through each sample as described in the next section. 

\subsection{Hypothesis Testing: Kolmogorov-Smirnov Statistical Test}

Hypothesis testing is a common form of statistical inference in which sample data are used to estimate unknown population parameters. In this study, we reverse this logic: we begin with an assumed population distribution and, through simulation, predict the number of samples needed to reconstruct its parameters, accounting for uncertainties due to sampling error and observation sequence (discussed in Sections 3.2 and 3.3 respectively). We apply a one-dimensional, two-sample, two-tailed Kolmogorov-Smirnov (KS) test to evaluate statistical differences between the sample subset and the model \citep{lehmannTestingStatisticalHypotheses2008}. Many use the KS test because it is non-parametric, requiring no specific distributional assumptions, and can handle the comparison of samples of varying sizes \citep{bergerKolmogorovSmirnovTest2014, masseyKolmogorovSmirnovTestGoodness1951}. The null hypothesis ($H_0$) states that both samples come from a population with the same unknown distribution, while the alternative hypothesis ($H_A$) states that the two samples come from different distributions. Although we retain these formal definitions, the test is not used to determine whether the samples share the same parent distribution, as this is already known. Instead, we are interested in using the test statistic as a measure of the statistical difference between the samples and the model.

This second level of sampling aims to determine the smallest sample size that is statistically indistinguishable from the model according to the KS test. In other words, we are interested in how much data is necessary before a sample of increasing size (ref. Figure 4) starts to resemble the population. The method involves comparing a sample subset to a sample from the N(0, 1) model, with the subset increasing one data point at a time until an error threshold is reached between the two. We define this threshold in the next section. The KS test measures the maximum vertical distance, D, between the cumulative distribution functions (CDFs) of the sample subset and the full sample. This distance is compared to a critical value, $D_{crit}$, visualized in Figure 4. At each iteration, when a new point is added to the subset, the KS test is recalculated, and D is compared to $D_{crit}$. When D falls below $D_{crit}$, the difference between the two CDFs is sufficiently small, suggesting that the sample, if representative of N(0,1), is likely drawn from the model. Note, increasing the number of evaluated samples, as we have done, mitigates sampling error associated with drawing samples that deviate from the underlying N(0,1) distribution. 

The final column of Figure 4 shows the PDFs of the sample subsets, estimated using kernel density estimation (KDE). As more samples are acquired in the subset (blue), the estimated PDF progressively converges toward the distribution of the assumed N(0, 1) model (red). Once the critical value is satisfied, the subset PDF is expected to match that of the model (red). Each KDE reconstruction of the model is recorded at every iteration (see Results section) to compare with the N(0, 1) model. This step evaluates the degree to which the reconstructions reproduce the underlying model distribution while minimizing the required sample size. 

\begin{figure}[ht!]
\includegraphics[width=\columnwidth]{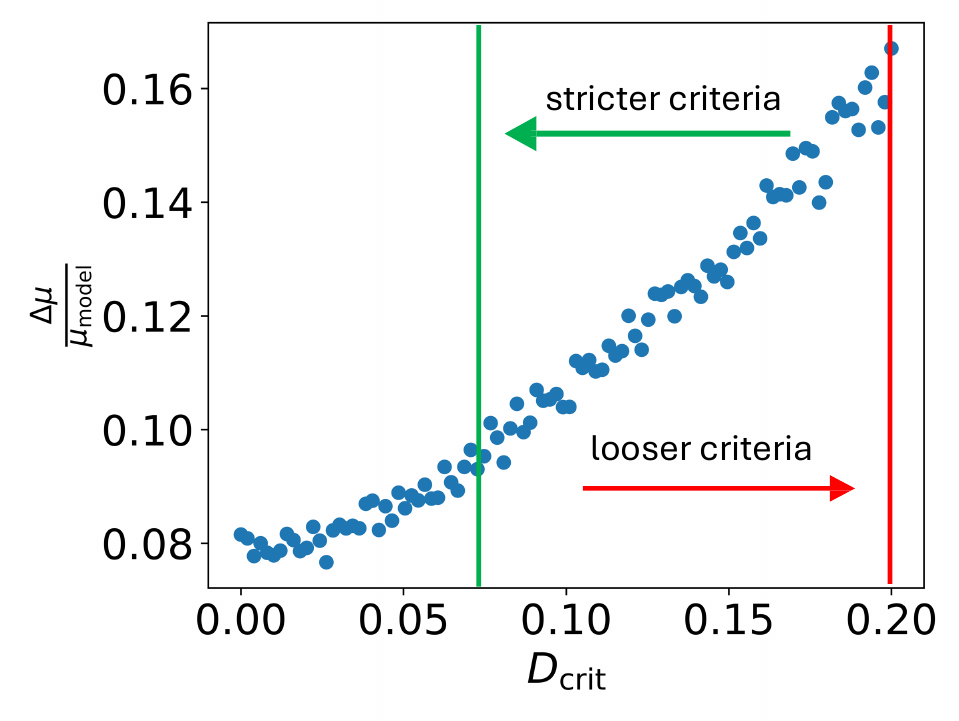}
\caption{\textbf{Sensitivity of reconstruction accuracy to the KS critical value ($D_{crit}$).} The x-axis shows $D_{crit}$ values; the y-axis shows the mean misfit between the reconstruction and the $N(0, 1)$ model. For each $D_{crit}$, 10,000 Monte Carlo trials were run, where subsets of samples from $N(0, 1)$ were compared to test for likeness and estimate the minimum sample size. Larger $D_{crit}$ values lead to higher misfit, consistent with expectations. Selected end-member criteria reflect both prior studies and the quality of the resulting model predictions.}
\end{figure}

\subsection{Defining Criticality} 

The KS test is typically used as a hypothesis test where the null hypothesis is rejected or fails to be rejected based on the value of a test statistic relative to a critical value. The critical value, a quantile of the null distribution, serves as a threshold that defines the rejection region of the null distribution, where a test statistic is unlikely to occur. In other words, critical values determine whether the results of a hypothesis test are statistically significant. 

In this study, we evaluate two critical values ($D_{crit}$) to represent varying error thresholds, and by extension, different observational outcomes. Figure 5 shows a sensitivity analysis from pilot simulations in which $D_{crit}$ was varied to assess how prediction error scales. Based on these results, we focus on two end-member cases: $D_{crit}$ = 0.2 and $D_{crit}$ = 0.075. The former represents a less stringent criterion, permitting larger discrepancies between the sample and model CDFs to capture broader distributional differences. This approach reduces the required sample size but increases the risk of Type I errors or false positives, making it most useful for early-stage analyses aimed at guiding subsequent observations. The latter represents a more stringent criterion, sensitive to smaller distributional differences. While this requires larger sample sizes, which come at a higher cost in terms of time and resources, it reduces the likelihood of Type I errors by requiring stronger evidence before rejecting the null hypothesis. In the results section, we provide confidence intervals for the number of observations required to satisfy each threshold and recover the parameters of the N(0,1) model.  

\section{Results} \label{subsec:results}

Using the methods described above, we evaluate the average sample size needed to determine a random sample from N(0,1) is statistically consistent with N(0,1). This controlled case provides a baseline for interpreting the evidence required to accept the null hypothesis under different observational constraints. These end-member tolerance levels—$D_{crit}$=0.20 (loose) and $D_{crit}$=0.075 (strict)—represent constraints given by poor vs. optimistic data yields. The next sections will provide results at two levels of analysis—first at the sequencing level for each tolerance level, and second, at the larger scale using Monte Carlo sampling to estimate how many $pCO_2$ observations are needed to constrain N(0,1) population parameters. 

\begin{figure*}[ht!]
\centering
\includegraphics[width=0.9\textwidth]{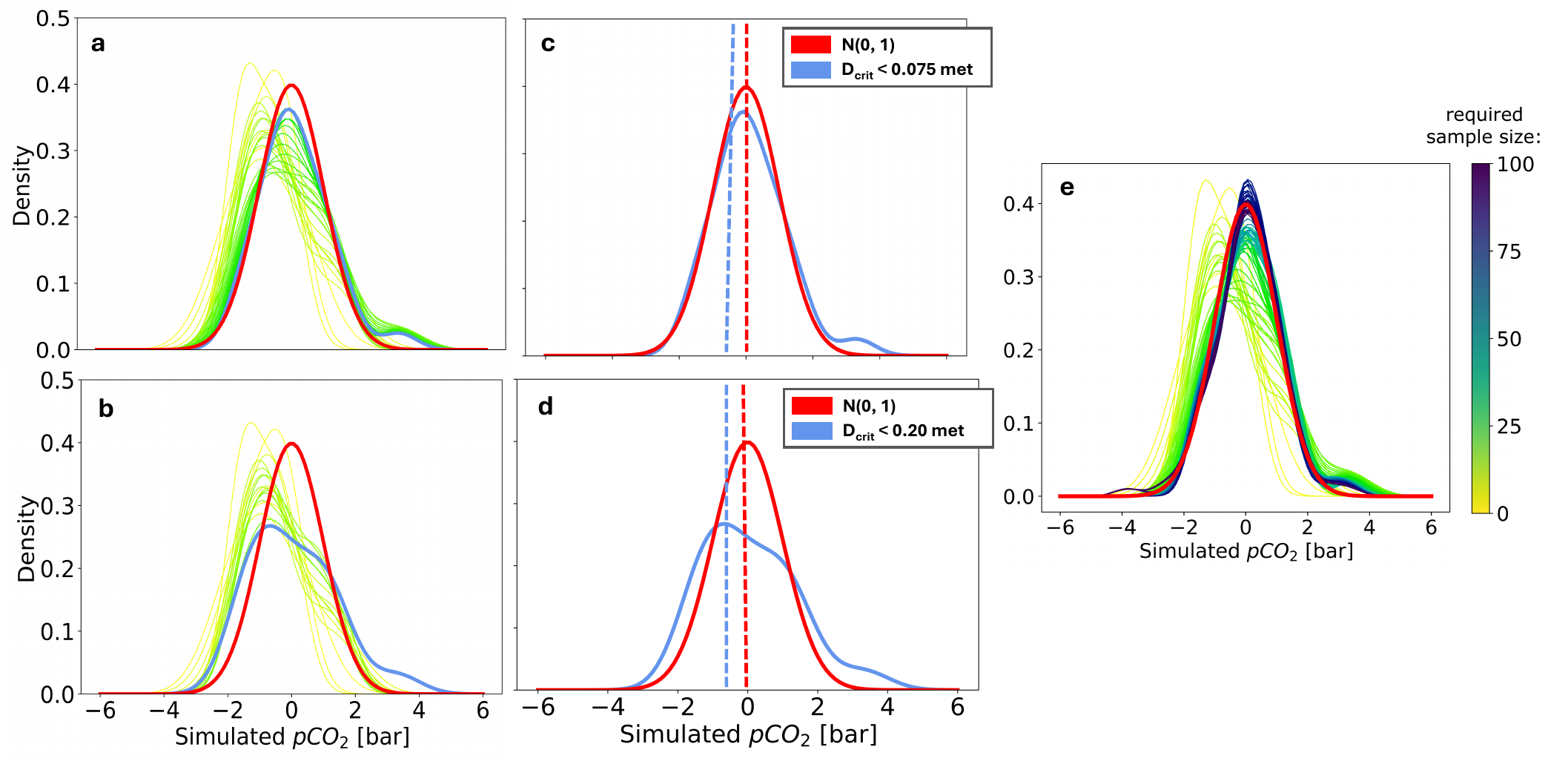}
\caption{\textbf{Example of one trial outcome where strict (a, c) and loose (b, d) error thresholds are used to reconstruct the model distribution, $N(0, 1)$.} The model distribution is red; the predicted distributions are in blue; and the color grade (ref. color bar in (e)) represents acquiring a data point in the sample subset. (a, c) shows the results of one trial where the reconstruction of the model distribution is estimated using the strict criterion $D_{crit}$ = 0.075. (b, d)  shows the same using the loose criterion $D_{crit}$ = 0.2.}
\end{figure*}

Figure 6 shows an example of how a sample distribution evolves under both high- and low- error tolerance. In each case, the sample PDF is estimated using KDE from discrete draws of the population. The red curve denotes the N(0, 1) model, while the color gradient tracks the KDE as more data points are sequentially added. For both tolerance levels, the same sample is used to highlight differences in data requirements and prediction outcome. The left column shows how distributions of increasing sample size change relative to the model; the middle column shows the final comparison at the point where the critical value is met. The distributions in Figure 6 represent a single illustrative sequence from N(0,1) and do not reflect population-level outcomes. Panel 6e shows the KDE reconstruction using all 100 points from the example sample (darker purple indicates closer agreement with the model). Because individual samples can be more or less representative of the population, this process is repeated 10,000 times using simple random sampling to capture variability from sampling error. The aggregated results are shown in Figure 7, where gray curves represent individual predictions (analogous to blue distributions in Figure 6), and the red curve again shows the N(0,1) model.  

\begin{figure*}[ht!]
\centering
\includegraphics[width=0.9\textwidth]{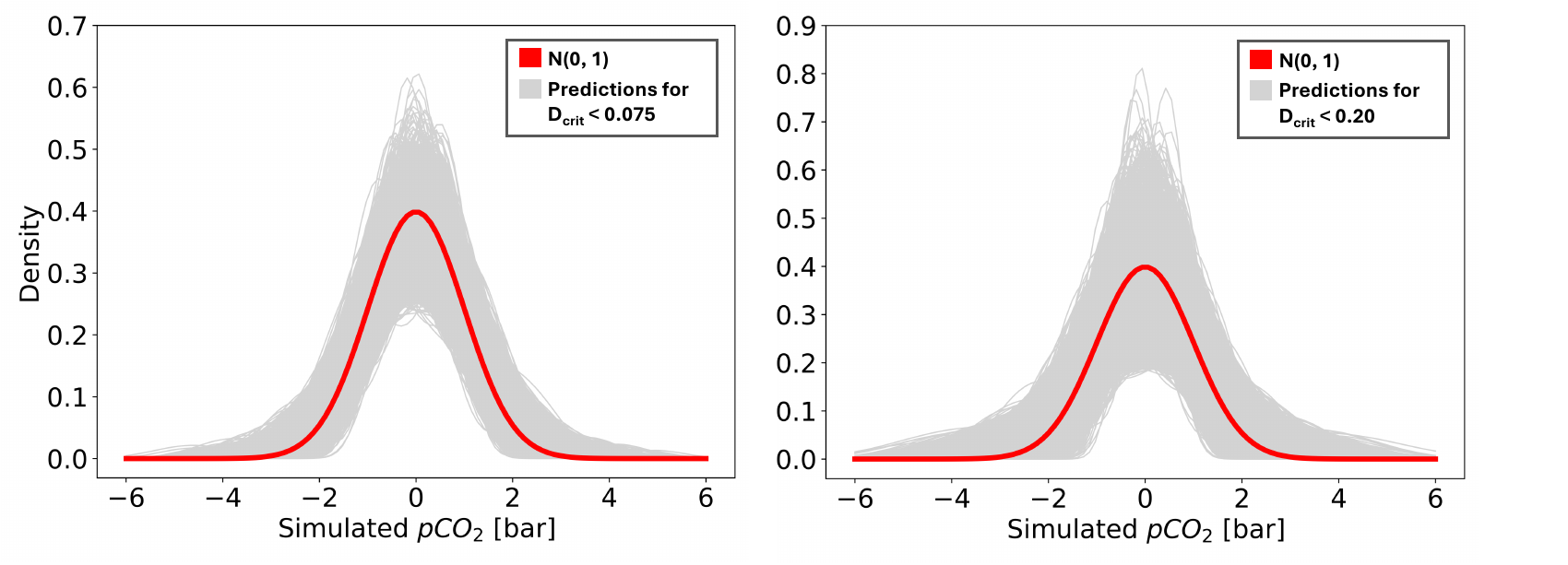}
\caption{\textbf{Reconstruction performance under different error thresholds.} The model distribution is shown in red; the gray lines show 10,000 predictions made using either the strict or loose cutoff criteria. In panel (a), a stricter $D_{crit}$ requires closer agreement with the model before confirming a match. The reconstructions are produced by [23, 74] (95\% CI) observations and reproduce the model well (ref. Figure 8 for uncertainty quantifications). Panel (b) shows the results of using a more lenient $D_{crit}$, which accepts a poorer fit. The results are less representative of the model, but require fewer observations [5, 29] (95\% CI).}
\end{figure*}

The results in Figure 7 show how the choice of threshold affects model fit. Under the stricter criterion, predictions more closely approximate the model (Fig. 7a), while the more permissive criterion, allows for broader, more variable fits (Fig. 7b). We quantitatively assess fit in Figure 8 using three complementary metrics: relative standard deviation (RSD), the log-ratio of standard deviations, and total variation distance (TVD).

\begin{figure*}[ht!]
\centering
\includegraphics[width=0.9\textwidth]{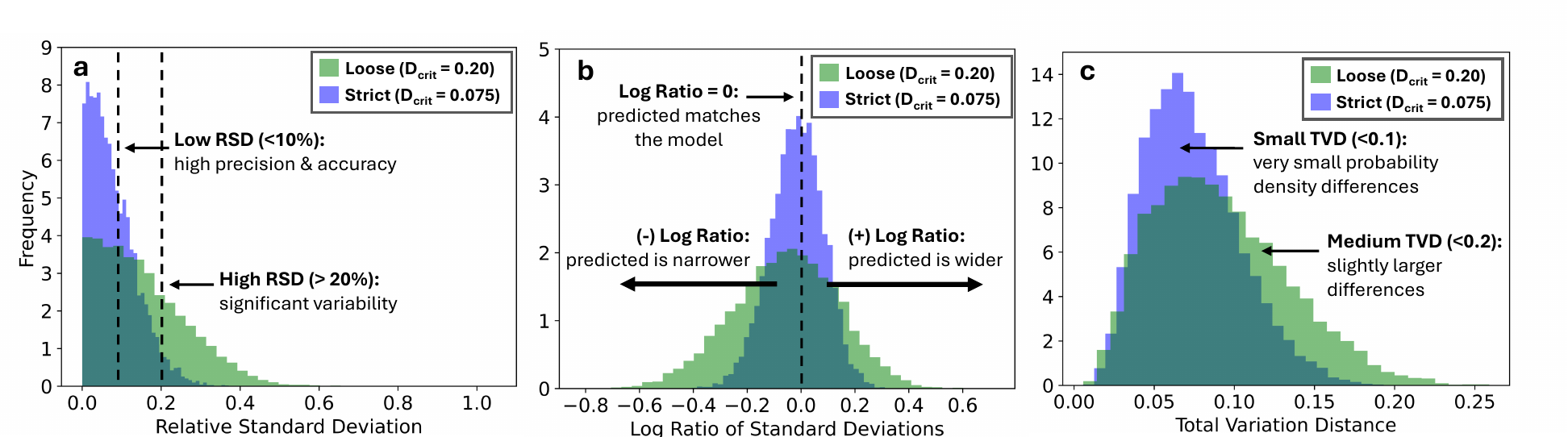}
\caption{\textbf{Accuracy of reconstructions under strict and loose $D_{crit}$ criteria.}  All panels compare the accuracy of the reconstructed distributions (from Figure 7) to the reference model $N(0, 1)$. Results obtained under the strict $D_{crit}$ threshold are shown in purple, while those from the looser threshold appear in green. (a) shows the relative standard deviation (RSD) between the model and each reconstruction. Values near zero indicate close agreement; larger deviations suggest poorer fits. (b) presents the log ratio of standard deviations. Positive values indicate reconstructions are broader than the model, negative values indicate narrower spreads, and values near zero imply a close match. (c) shows the total variation distance (TVD) between each reconstruction and the model. Lower TVD values reflect smaller differences in probability density and a better overall fit. These metrics demonstrate that the stricter $D_{crit}$ yields more accurate reconstructions, whereas the looser criterion yields worse fits but requires fewer observations.}
\end{figure*}

RSD gives a unitless, normalized comparison between the spread of each prediction and that of the reference model:
\[
\mathrm{RSD} = \frac{|\sigma_{\text{predicted}} - \sigma_{\text{model}}|}{\sigma_{\text{model}}}
\]
Figure 8a plots the distribution of RSD values found under both the strict and loose criteria. Values near 0 indicate close agreement with the model, whereas larger departures from 0 indicate broader or narrower predicted spreads. As expected, the stricter criterion generates values tightly clustered around 0, while the looser criterion admits a wider range of deviations. 

The log-ratio of standard deviations highlights the direction as well as the magnitude of any spread mismatch. 
\[
\ log(\frac{\sigma_{\text{predicted}}}{\sigma_{\text{model}}})
\]
In Figure 8b, a log-ratio of 0 signifies a perfect match. Positive values indicate that the prediction is broader than the model; negative values indicate it is narrower. Both criteria generate symmetric tails, but the strict criterion concentrates mass more near 0, confirming its tighter control on model-prediction differences. 

Lastly, TVD measures the difference in probability mass between a prediction and the model:
\[
\mathrm{TVD(F, G)} = \frac{1}{2} \int |f(x)-g(x)|\,dx
\]
Figure 8c shows that TVD values under the strict criterion cluster closer to 0, implying minimal redistribution of probability density. By contrast, the loose criterion permits noticeably larger TVDs, reflecting greater overall divergence. 

Together, these metrics demonstrate that the strict criterion more closely predicts the model’s statistical structure, whereas the loose criterion permits substantially more variation in both spread and distribution shape. As Figure 7 shows, this difference is also tied to how much data is needed: achieving a good fit under the strict criterion requires a larger sample size, whereas the loose criterion reaches acceptable fits with fewer data points but at the cost of precision.  

\begin{figure}[ht!]
\includegraphics[width=0.9\columnwidth]{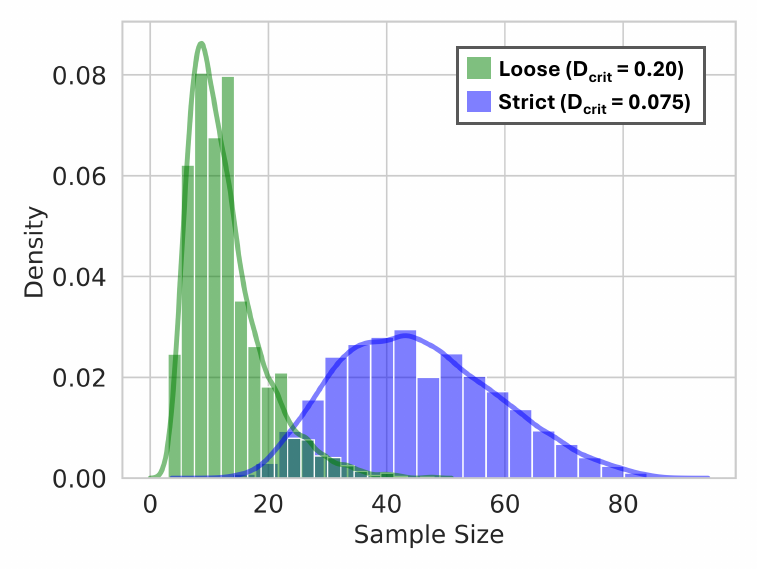}
\caption{\textbf{Required sample sizes for detecting distribution similarity under different error thresholds.} This figure shows the range in sample sizes needed to obtain the fits shown in Figure 7. The x-axis represents the number of observations needed to confirm similarity to the model distribution using the two end-member criteria. The results produced under the strict criterion generally required more observations than the looser one. Notably, the distributions of required sample sizes are asymmetric—this influences how we report uncertainty, as the standard symmetric confidence intervals ($2\sigma$) are no longer appropriate.}
\end{figure}

\begin{table}[h]
    \centering
    \label{tab:your-label}
    \includegraphics[width=\columnwidth]{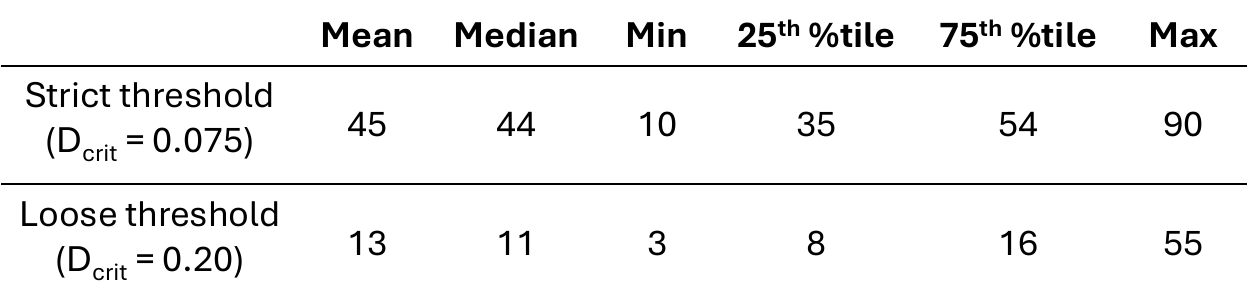}
    \caption{\textbf{Summary statistics for the required sample sizes found under the different error thresholds.}  For results under the strict and loose criteria, we report summary statistics for the required sample sizes. Because the distributions are asymmetric, the mean is pulled away from the center of the data, leading to a misleading representation of the typical value. Thus, we report the median along with the interquartile range for a more accurate and robust representation of the central tendency and spread. The minimum and maximum are also reported to give one an idea of the magnitude of spread and the upper and lowermost bounds for planet observations.}
\end{table}

Figure 9 shows the distributions of the minimum sample sizes required to achieve the results from Figures 7a and 7b. As both distributions are right-skewed, we report uncertainty using percentile-based confidence intervals rather than standard $\pm2\sigma$ bounds. This approach is commonly used in frequentist Monte Carlo analyses to account for asymmetry in the distribution of the estimator. Using the 2.5th and 97th percentiles, we calculate the 95\% confidence interval for the stricter criterion as [23, 74] samples, and for the permissive criterion [5, 29] samples. These intervals represent the range in which we would expect the true parameter value to fall 95\% of the time, if we were to repeat this simulation many times under the same conditions. Summary statistics are provided in Table 1 for further context. 

\begin{figure*}[ht!]
\includegraphics[width=0.9\textwidth]{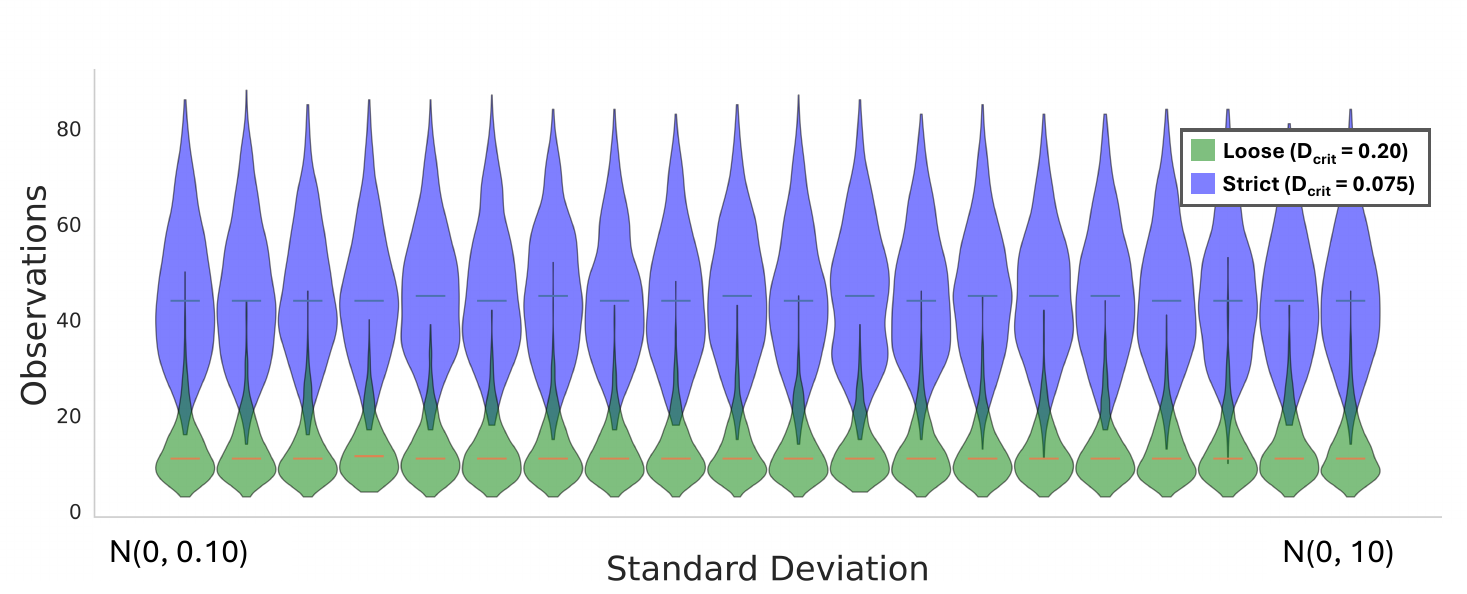}
\caption{\textbf{Influence of normal distribution parameters on required sample size.} Each violin represents sample size estimates under a slightly different normal distribution assumption, with purple and green indicating strict and loose criteria, respectively. The bars within each violin show the median estimate. While varying the mean does not affect the required sample size and is not particularly of interest here, we are interested in changes to the standard deviation, which relates to feedback strength. However, results show that even as standard deviation increases, the required sample size remains largely unchanged, indicating that a similar amount of data can constrain models with broader variability assumptions.}
\end{figure*}

The results in Figure 9 for a model assumption of N(0, 1) also hold across other normal distributions with smaller and larger standard deviations. Figure 10 presents outcomes for initial population distributions ranging from N(0, 0.1) to N(0, 10). The green violins correspond to tests with higher error tolerances, which require smaller sample sizes but yield less accurate estimates. In contrast, the purple violins correspond to lower error tolerance, which need larger sample sizes to achieve better fit. Interestingly, the stricter tolerance produces wider confidence intervals—contrary to the expectation that increasing sample size improves precision and narrow the confidence interval. This pattern, also apparent in the N(0, 1) case, reflects the sensitivity of the KS test under more stringent critical value. Because the test depends on the maximum difference between cumulative distributions, even small discrepancies are amplified at larger sample sizes, leading to greater apparent variability.

\section{Discussion} \label{subsec:discuss}

Although it is a prime motivator, characterizing the atmospheric properties of terrestrial exoplanets goes beyond just searching for biosignatures (indicators that the presence of life may have altered a planet’s atmosphere). It can provide a means for testing planetary hypotheses at a level that is not possible using only observations from our own solar system. A hypothesis we have focused on in this paper is that of climate buffering. Long-term climate stability on Earth is thought to be maintained by a negative feedback cycle that regulates greenhouse gas concentrations, and, by extension, surface temperature. Whether such buffering is a characteristic of terrestrial planets that share commonalities with Earth (e.g. bulk composition, solar proximity) is unknown. Future observations from these planets can potentially provide support that climate buffering extends beyond Earth or that it may be a characteristic unique to Earth. Either answer would be valuable information. 

If climate feedback is present on terrestrial planets, within orbital windows around similar stars, then this should produce detectable patterns in the probability distributions of key variables like atmospheric $CO_2$. The presence of feedback means that atmospheric $CO_2$ will tend to resist long-term shifts and return toward a central tendency (see Figure 3). For the Earth, a planet we have historical records for, the time variation of atmospheric $CO_2$ can be used to determine whether greenhouse gas buffering has been operative through time. A similar test is possible for a population of Earth-like planets even though historical records cannot be extracted. Even if Earth-like planets are all observed at the same absolute time, they will be associated with different evolutionary times. Those differences will lead to differences in atmospheric properties, but if buffering feedback(s) are at work, we would expect to see a peaked distribution when a number of terrestrial planet observations are compiled. This is a testable prediction that motivates a focus point of our study: what is the minimum number of observations needed to constrain a distribution of $pCO_2$ to determine if it is consistent with the hypothesis of climate buffering? 

The question of how many observations are needed to adequately constrain the distribution of a characteristic in a population is a common one in statistical analysis. In most applications, this question is posed before data collection begins, allowing researchers to design studies with sufficient statistical power to address key hypotheses, like, for example, determining the number of clinical trials in a medical study or the number of surveys needed to assess population trends or preferences. However, for terrestrial exoplanets, the nature of data acquisition is the reverse of this logic. Observations will arrive opportunistically as the community is able to obtain them, and we have no \textit{a priori} knowledge of how many will ultimately be available. Moreover, target selection is likely to be driven by practical constraints, such as which planets are easiest to observe and characterize within reasonable time and position windows. For these reasons, although the core statistical methods used in our methodology are standard, the way we combine them is unique to the problem at hand. In particular, we account for uncertainties in the order in which observations from a given population may come to us—something not known to us beforehand—and we explore different strictness of critical values to represent two contrasting data availability scenarios.

Short of sampling every member of a given planetary population, there will always be uncertainty about whether enough planets have been observed to constrain the distribution of a given property (e.g. atmospheric greenhouse gas concentrations). In our methodology, the chosen critical values reflect the level of statistical risk that the community is willing to accept in this regard. Ideally, this risk is minimized by collecting more data, but practical limits—like the unknown total number of future observations and cost limitations—make such minimization challenging. Our approach was to first consider a higher error tolerance (or higher critical value), where we determine the sample size necessary to obtain a model fit that meets the ``loose" criterion. In this case, the criterion is more likely to be met in smaller sample sizes, due to the higher error tolerance. Using smaller sample sizes for statistical inference carries an added risk of false positives, but, in some cases, can provide preliminary support for (or evidence against) a distribution that, in our case, is consistent with climate buffering. Inferences made with this ``higher-risk" sample size can then be used to guide an early, coarse-grained assessment that can inform future target selection and observational strategy. Our stricter, lower error tolerance (or lower critical value) then represents a higher-return data scenario where access to larger sample sizes allows for better model fits, and there is a ``lower risk" of false positives and negatives. This standard is chosen because the better distribution fit allows for better constraints on statistical indicators— such as variance, skewness, kurtosis, and multimodality—that allow deeper investigation of feedback strength and mechanism.  

One central trade-off presented here is between sample size and statistical confidence. While larger samples yield stronger inferences, achieving them is not feasible given current telescope time and instrument constraints, something that will ideally be addressed by future flagship missions, the Nancy Grace Roman Space Telescope and HWO. Recognizing this difficulty, we structured our method to accommodate discrete luminosity bins, allowing for preliminary inference even with sparse data. Ideally, sufficient observations would be available to support higher-level inference or inform critical decisions, but the framework remains adaptable when that is not the case. This design also accommodates the possibility that distinct climate feedbacks operate at different luminosities, an outcome that, if observed, would contradict the HZ.

The presented approach can also be extended to test the classical HZ hypothesis. HZ theory posits that the carbonate-silicate weathering feedback drives atmospheric $CO_2$ of Earth-like planets to increase with orbital distance, modulating surface temperatures as stellar flux declines. Using our approach, we estimate the number of exoplanet observations required to detect the presence—and in some cases, the strength—of such a feedback across three discrete stellar luminosity windows (e.g., 0.9, 1.0, 1.1 $S/S\oplus$). Our results suggest that, optimistically, 90 and conservatively up to 222 observations (95\% CI) at minimum would be needed to statistically confirm or reject the HZ hypothesis. These numbers provide concrete targets for exoplanet surveys aiming to evaluate planetary climate stability. Our sample size estimates are larger than those of prior studies, some of which rely on model-dependent assumptions or use traditional null hypothesis significance testing, where the outcome is limited to rejecting or not rejecting the null, and the p-value does not quantify the degree of support for either hypothesis. Accounting for uncertainty in the sequence of observations also leads to larger numbers, especially in the conservative case as the increased number of possible observation sequences introduces greater variability in when the critical evidence may appear. 

In addition to error tolerance, our results highlight the importance of uncertainties arising from the order in which observations will come to the community. This type of uncertainty creates room for varied interpretations of whether the number of observations available at any given time are sufficient—given decisions about the statistical risk we are willing to accept—to assess consistency between the data and our hypothesis. We deliberately avoid prescribing a specific error tolerance that should or should not be used, as that is ultimately a community decision, and, given the nature of exoplanet observations, is something that will likely need to be adjusted as new data become available. In practice, using exoplanet observations to test any hypothesis will involve risk-cost-benefit decisions. While this study outlines some risks associated with observational uncertainty (sequencing and sample size), it does not address observational cost. And ``benefit" depends on which scientific questions the community wishes to prioritize (e.g., arguably the priority question for Earth-like planets at present is one of biosignature detection—what the next level questions are is a matter the community will decide on). 

Instrument and measurement uncertainty will likely increase the observational requirements estimated in this study. Retrieving absolute atmospheric $CO_2$ abundances from exoplanet spectra depends not only on signal quality but also on the calibration accuracy of the telescope and detectors, as well as on atmospheric-retrieval algorithms that introduce their own biases and degeneracies. These issues are better documented for larger gaseous planets \citep{greeneCHARACTERIZINGTRANSITINGEXOPLANET2016, kemptonTransitingExoplanetAtmospheres2024, rocchettoEXPLORINGBIASESATMOSPHERIC2016}, but remain less explored for smaller Earth-sized planets, where weaker signals exacerbate noise and systematic errors. As both instrument limitations and retrieval uncertainties become better characterized, they can be incorporated into these types of frameworks to refine sample size requirements for population-level tests of terrestrial planet characteristics. 

Future work will explore how our observational needs shift under different assumptions about population parameters. Being in the HZ is a necessary but insufficient condition for a planet to have liquid water at its surface. The geological history of the Earth provides evidence for sustained snowball and moist-greenhouse states in the past, despite remaining in the HZ. If multiple stable climate states exist for the Earth, then it is possible that multimodal $pCO_2$ could be observed for Earth-like planets at comparable solar distances within the habitable zone \citep{muranteClimateBistabilityEarthlike2020, menouClimateStabilityHabitable2015, boschiBistabilityClimateHabitable2013, lucariniHabitabilityMultistabilityEarthlike2013, grahamCO2OceanBistability2022}. This could track with the Earth-Venus dichotomy in our own solar system, where both planets are similar in terms of size, bulk compositions, and solar distances but maintain very different climate states. Future work will address this possibility and how the observational need will change to test a multimodal hypothesis. 

\section{Conclusion} \label{subsec:conc}

As it stands, we still lack enough atmospheric data for terrestrial exoplanets to evaluate most habitability hypotheses. However, a growing number of recent studies have proposed empirically testable hypotheses, marking progress toward framing our habitability questions in ways that can be directly evaluated \citep{affholderInteriorConvectionRegime2025, grahamThermodynamicEnergeticLimits2020, lehmerCarbonatesilicateCyclePredictions2020b, beanStatisticalComparativePlanetology2017a}. By estimating required sample sizes or detection thresholds, such studies inform mission planners whether a hypothesis is testable with current or near-term instruments, or if it will require future flagship missions. A natural next step in this approach is to develop adaptive frameworks that respond to new observations, where early results guide target selection, and subsequent rounds of data acquisition refine our models and understanding. In this work, we extend previous efforts to test for an active carbonate silicate weathering cycle on HZ planets, by 1) minimizing assumptions about the processes that regulate atmospheric $CO_2$; 2) basing our test on error tolerance or the ``statistical risk" one is willing to take; and 3) providing the sample sizes needed to meet those error tolerances. Our findings suggest that between 30 to 74 observations may be sufficient to detect a $CO_2$-buffering feedback, with smaller samples carrying a higher risk of false positives and larger samples yielding the closer fits needed to estimate feedback strength. We extend this framework to estimate the sample size necessary to test the HZ hypothesis, finding that a minimum of 90 observations would be needed to confidently detect feedback behavior.

\bibliographystyle{aasjournalv7}

\end{document}